\documentclass[twocolumn,showpacs,amsmath,amssymb,prd]{revtex4}
\newcommand{\beq}{\begin{equation}}
\newcommand{\eeq}{\end{equation}}
\newcommand{\beqn}{\begin{eqnarray}}
\newcommand{\eeqn}{\end{eqnarray}}

\newcommand{\ve}[1]{\mbox{\boldmath $#1$}}

\usepackage{graphicx}
\usepackage{dcolumn}
\usepackage{bm}
\usepackage{epsf}

\begin{document}

\title{Excitation Of MHD Modes With Gravitational Waves: \\
A Testbed For Numerical Codes}

\author{Matthew D. Duez}

\author{Yuk Tung Liu}

\author{Stuart L.\ Shapiro}

\altaffiliation{Also Department of Astronomy \& NCSA, University of Illinois
at Urbana-Champaign, Urbana, IL 61801}

\author{Branson C. Stephens}

\affiliation{Department of Physics, University of Illinois at
Urbana-Champaign, Urbana, IL~61801}

\begin{abstract}
We consider a gravitational wave oscillating in an initially 
homogeneous, magnetized
fluid. The fluid is perfectly conducting and isentropic, and the
magnetic field is initially uniform. We find analytic solutions
for the case in which the gravitational wave is linear and
unaffected by the background fluid and field. Our solutions
show how gravitational waves can excite three 
magnetohydrodynamic (MHD)
modes in the fluid: Alfv\'en waves, and both fast and slow
magnetosonic waves. Our analytic solutions are particularly useful
for testing numerical codes designed to treat general relativistic
MHD in {\em dynamical} spacetimes, as we demonstrate in a
companion paper.
\end{abstract}


\maketitle
                                                                                     
\section{Introduction}
\label{intro}

Relativistic magnetohydynamics (MHD) plays an important role in high-energy 
astrophysics. Quasars, active galactic nuclei, X-ray binaries, and 
gamma-ray bursts (GRBs) are thought to be powered by black holes or neutron 
stars. In most cases, relativistic plasmas and strong magnetic fields 
are believed to be involved 
in the extraction of energy from the central objects. 

In the past decade, numerical codes have been developed to simulate 
MHD fluids in general 
relativity~\cite{yokosawa93,koide99,koide00,devilliers03,gammie03,komissarov04}. 
However, most of the codes to date have assumed that the background 
metric is stationary and fixed. This 
assumption is valid in many astrophysical scenarios and it greatly 
simplifies the numerical calculations. Such is the case, for example, 
for gas accretion onto Kerr black holes whenever the self-gravity 
of the fluid can be ignored. However, the spacetime is not 
stationary in many other cases. In the `collapsar' 
model of GRBs~\cite{macfadyen99}, in relativistic stars undergoing 
magnetic braking of differential rotation~\cite{mag_braking}, 
and in the catastrophic collapse 
of hypermassive neutron stars~\cite{hypermassive},
the spacetime is highly dynamical, and codes based on a fixed, stationary
background metric are not able to simulate these scenarios.

We have recently developed the first general relativistic (GR) 
MHD code~\cite{fn0} 
that evolves the spacetime metric, together with 
the fluid, by integrating the coupled Einstein-Maxwell-MHD equations 
(the ``GR MHD equations'' for short) in 3+1 dimensions 
without approximation~\cite{duez05} (hereafter, Paper~I). 
We expect that similar codes will be developed by other groups in 
the near future. To verify our code, it is important that they  
pass some nontrivial test problems involving {\em dynamical} spacetimes. 
There already exists a suite of standard tests for GR MHD codes in 
Minkowski spacetime~\cite{komissarov99,gammie03,devilliers03}, 
as well as tests in stationary curved 
spacetimes~\cite{koide99,gammie03,devilliers03}. 
However, very few nontrivial problems have been proposed to test 
a GR MHD code in dynamical spacetimes. The evolution of the MHD interior 
and vacuum exterior magnetic field in the background spacetime of a 
spherical, homogeneous dust ball undergoing collapse (``magnetized 
Oppenheimer-Synder collapse'') has been solved and provides one 
such test problem~\cite{baumgarte03}. 
The purpose of this paper is to formulate another such test problem, 
one which is analytic and very straightforward to implement in a pure 
MHD environment. 
We consider a gravitational wave oscillating in an initially homogeneous
fluid immersed in a uniform magnetic field. The gravitational wave 
will, in general, excite three wave modes in the fluid: 
Alfv\'en waves, and fast and slow magnetosonic 
waves. Assuming that the gravitational wave is 
weak, we linearize the GR MHD equations and find analytic solutions. 
This problem has been studied previously by several authors in 
other contexts
(see~\cite{psvk01,moortgat03,moortgat04,kallberg04} and references therein).
However, the equations in most of those papers are written in the orthonormal 
tetrad frame~\cite{ft0}, whereas most of the GR MHD codes today evolve components 
of the MHD variables in a coordinate basis. The purpose 
of this paper is to derive the linearized solution 
simply and present it in a form 
that can be used to compare directly with the numerical results.
We also consider a different physical scenario whose solution is 
more suitable for numerical code tests than the solutions found 
in previous studies of similar problems.
We recently have performed numerical simulations for the test problem 
formulated in this paper with our new GR MHD code and found good 
agreement with our analytic
solutions. The details of our numerical code and simulations
are reported in Paper~I.

The structure of this paper is as follows. We briefly summarize the GR 
MHD equations in Section~\ref{sec:GRMHDeqs}. In 
Section~\ref{sec:linearized-eqs}, 
we derive the linearized equations. Since most numerical codes solve 
the fully nonlinear equations, it is important to know in what regime 
the linearized solution applies. Therefore, we list the 
necessary conditions for the linearization 
procedure to be valid. We then solve the linearized equations in 
Section~\ref{sec:solution} and summarize our results in 
Section~\ref{sec:summary}.

\section{General relativitic MHD equations}
\label{sec:GRMHDeqs}

Throughout, we adopt geometrized units in which $c=G=1$. 
Greek (spacetime) indices run from 0 to 3, while Latin (spatial) indices 
run from 1 to 3. The signature of the metric is $(- + + +)$.

The law of baryon number conservation gives rise to the 
continuity equation $\nabla_{\mu} (\rho_0 u^{\mu})=0$, where 
$\rho_0$ is the rest-mass density and $u^{\mu}$ is the 4-velocity. 
The continuity equation can be written as 
\beq
  \partial_t \rho_* + \partial_i (\rho_* v^i) = 0 \ ,
\label{eq:con}
\eeq
where $\rho_* = \sqrt{-g}\, \rho_0 u^t$, $v^i = u^i/u^t$, and 
$g$ is the determinant of the metric $g_{\mu \nu}$. Conservation of 
energy-momentum $\nabla_{\mu} T^{\mu}{}_{\nu}=0$ can be written as 
\beq
  \partial_t (\sqrt{-g}\, T^t{}_{\nu}) + \partial_j 
(\sqrt{-g}\, T^j{}_{\nu}) = \frac{1}{2} \sqrt{-g}\, T^{\alpha \beta} 
g_{\alpha \beta, \nu} \ .
\label{eq:energy-momentum}
\eeq
The time component of this equation gives the energy equation, and 
the spatial components give the momentum equations.  

The stress-energy tensor of a magnetized fluid is given by 
\beq
  T^{\mu \nu} = \rho_0 h u^{\mu} u^{\nu} + P g^{\mu \nu} 
+  T^{\mu \nu}_{\rm em} \ ,
\eeq
where $P$ is the gas pressure. The specific 
enthalpy $h$ is related to the specific internal energy $\epsilon$ by 
$h=1+\epsilon + P/\rho_0$. We adopt a $\Gamma$-law equation of state 
$P=(\Gamma-1)\rho_0 \epsilon$.
The stress-energy tensor for the electromagnetic field is given by 
\beq
  T^{\mu \nu}_{\rm em}= \frac{1}{4\pi} \left( F^{\mu \lambda} 
F^{\nu}{}_{\lambda}
- \frac{1}{4} g^{\mu \nu} F_{\alpha \beta} F^{\alpha \beta} \right)  \ ,
\eeq
where $F^{\mu \nu}$ is the electromagnetic field tensor. 
In the ideal MHD regime, where the electric field vanishes 
in fluid's frame, it is possible to write $F^{\mu \nu}$ in terms of 
a magnetic vector field $b^{\mu}$ as follows:
\beqn
  F^{\mu \nu} &=& \sqrt{4\pi} \epsilon^{\mu \nu \kappa \lambda} 
u_{\kappa} b_{\lambda} \ , \\
  b^{\mu} &=& \frac{1}{2\sqrt{4\pi}} \epsilon^{\mu \nu \kappa \lambda} u_{\nu}
F_{\lambda \kappa} = \frac{B_{(u)}^{\mu}}{\sqrt{4\pi}} \ ,
\eeqn
where $\epsilon^{\mu \nu \kappa \lambda}$ is the Levi-Civita tensor,
and $B^{\mu}_{(u)}$ is 
the magnetic field measured by an observer comoving with the fluid.
In terms of $b^{\mu}$, the EM field stress-energy tensor becomes 
\beq
  T^{\mu \nu}_{\rm em} = b^2 u^{\mu} u^{\nu} + \frac{1}{2} b^2 g^{\mu \nu}
 - b^{\mu} b^{\nu} \ ,
\eeq
where $b^2=b^{\mu} b_{\mu}$.

The evolution of the EM fields is governed by the source-free part of 
the Maxwell equations 
$F_{\mu \nu , \lambda} + F_{\lambda \mu, \nu} + F_{\nu \lambda , \mu}=0$. 
This equation is equivalent to 
\beq
\nabla_{\mu} F^{* \mu \nu} = 0 \ ,
\label{divFstar}
\eeq
where the dual tensor $F^{* \mu \nu}$ is defined as 
\beq
   F^*_{\mu \nu} = \frac{1}{2} \epsilon_{\mu \nu \kappa \lambda}
F^{\kappa \lambda} \ .
\eeq
We also introduce the vector $B^{\mu} = n_{\nu} F^{* \mu \nu} 
=\alpha F^{*\mu t}$, 
where $n^{\mu}$ is the time-like normal vector orthogonal to the 
$t$=constant hypersurface which satisfies $n^{\mu} n_{\mu}=-1$, 
and where $\alpha$ is the 
lapse function. The 
vector $B^{\mu}$ is the magnetic field measured by a normal observer 
moving with 4-velocity $n^{\mu}$ (see Paper~I), and is purely spatial 
($n_{\mu} B^{\mu}=0=B^t$). The 
vector $b^{\mu}$ is related to $B^{\mu}$ by (see Paper~I for a derivation)
\beq
  b^{\mu} = -\frac{P^{\mu}{}_{\nu} B^{\nu}}{\sqrt{4\pi}\, n_{\nu} u^{\nu}} \ ,
\label{eq:bB}
\eeq
where $P_{\mu \nu} = g_{\mu \nu} + u_{\mu} u_{\nu}$. 

In terms of $B^{\mu}$, the time component of Eq.~(\ref{divFstar}) 
gives rise to the magnetic constraint equation 
\beq
  \partial_i (\sqrt{\gamma} B^i ) = 0 \ ,
\label{divB}
\eeq
where $\gamma$ is the determinant of the 3-dimensional spatial metric 
$\gamma_{ij}=g_{ij}$. The spatial component of Eq.~(\ref{divFstar}) 
gives the magnetic induction equation 
\beq
  \partial_t (\sqrt{\gamma} B^i) + \partial_j 
[ \sqrt{\gamma} (v^j B^i - v^i B^j) ] = 0 \ .
\label{eq:induction}
\eeq

In summary, the GR MHD equations for the fluid and electromagnetic fields 
are the continuity equation~(\ref{eq:con}), 
the energy-momentum equations~(\ref{eq:energy-momentum}), the magnetic 
constraint equation~(\ref{divB}), and the induction 
equation~(\ref{eq:induction}). The MHD stress-energy tensor is 
\beq
  T^{\mu \nu} = (\rho_0 h + b^2) u^{\mu} u^{\nu} + \left( P + \frac{b^2}{2} 
\right) g^{\mu \nu} - b^{\mu} b^{\nu} \ .
\label{Tab:ba}
\eeq
The vector $b^{\mu}$ is calculated from $B^i$ using Eq.~(\ref{eq:bB}). 

\section{Linearized equations in the presence of gravitational waves} 
\label{sec:linearized-eqs}

Consider a gravitational wave $h_{\mu \nu}$ oscillating inside 
perfectly conducting fluid immersed in a magnetic field. In the 
absence of the wave, the fluid is homogeneous and static and the 
magnetic field is uniform. The spacetime metric can be decomposed as
\beq
  g_{\mu \nu} = \eta_{\mu \nu} + h_{\mu \nu} + h^{\rm matter}_{\mu \nu} \ ,
\label{eq:metric}
\eeq
where $\eta_{\mu \nu}$ is the Minkowski metric and $h^{\rm matter}_{\mu \nu}$
is the metric perturbation generated by the magnetized fluid. We 
assume that the wave is linear 
\beq
|h_{\mu \nu}| \ll |\eta_{\mu \nu}| \ .
\label{cond:linear}
\eeq
We wish to ignore $h^{\rm matter}_{\mu \nu}$ in our computation, 
so we require the condition
\beq
  |h^{\rm matter}_{\mu \nu}| \ll |h_{\mu \nu}| 
\label{cond:hmatter}
\eeq
in the region of interest.

The magnitude of $h^{\rm matter}_{\mu \nu}$ can be estimated as follows. 
In the absence of gravitational waves, we set up Riemann normal 
coordinates (see Ref.~\cite{mtw}, Section 11.6) at a given point inside 
the fluid. The metric near the point is then given by 
\beq
  g_{\mu \nu} = \eta_{\mu \nu} - \frac{1}{6} (R_{\alpha \mu \beta \nu} 
+ R_{\alpha \nu \beta \mu}) x^{\alpha} x^{\beta}  \ ,
\eeq
where $R_{\alpha \mu \beta \nu}$ is the Riemann tensor. Hence we have 
\beq
  h^{\rm matter}_{\mu \nu} =  - \frac{1}{6} (R_{\alpha \mu \beta \nu} 
+ R_{\alpha \nu \beta \mu}) x^{\alpha} x^{\beta} 
\eeq
and $|h^{\rm matter}_{\mu \nu}| \sim \mathcal{R} r^2$, where 
$\mathcal{R}$ is the magnitude of the Rieman tensor. 
From Einstein's equation, $\mathcal{R}$ is of 
order $|T_{\mu \nu}|$. Hence, we have $|h^{\rm matter}_{\mu \nu}| \sim 
|T_{\mu \nu}| r^2$. We consider the regime where $h^{\rm matter}_{\mu \nu} 
\ll \eta_{\mu \nu}$, which implies that
\beq
  r \ll 1/\sqrt{|T_{\mu \nu}|} \ .
\label{cond:r}
\eeq
Equivalently, the region we consider should be much less than the 
radius of curvature.

Condition~(\ref{cond:hmatter}) can be written as
\beq
  |T_{\mu \nu}| \ll |h_{\mu \nu}|/L^2 \ ,
\eeq
where $L$ is the size of the region we consider (i.e., the computational 
domain). Note that this condition is more stringent than Eq.~(\ref{cond:r}) 
since $|h_{\mu \nu}| \ll |\eta_{\mu \nu}|$.
We will consider the situation where $L$ is of 
the order of the wavelength of the gravitational wave. Hence we have
$|T_{\mu \nu}| \ll k^2 |h_{\mu \nu}|$, where $k$ is the wave number of
the gravitational wave. Let $\rho_0$ and $P_0$ be the rest-mass density
and pressure of the unperturbed fluid and $B_0$ be the magnitude of
the unperturbed magnetic field. Then the
magnitude of $T_{\mu \nu}$ is of order $|T_{\mu \nu}| \sim
\rho_0(1+\epsilon_0)+P_0 + B_0^2/4\pi$, where $\epsilon_0$ is
the specific internal energy of the unperturbed fluid.
Condition~(\ref{cond:hmatter}) can be written as
\beq
  {\cal E} \ll k^2 h_0 \ ,
\label{cond:matter}
\eeq
where $h_0=|h_{\mu \nu}|$ is the magnitude of $h_{\mu \nu}$, and
\beq
  {\cal E} = \rho_0(1+\epsilon_0)+P_0 + \frac{B_0^2}{4\pi} \ .
\label{def:w}
\eeq
Our linearized solution is only valid for a time much 
less than the dynamical collapse time of the unperturbed fluid; 
otherwise the background spacetime will evolve appreciably due to 
self-gravity:
\beq
 t \ll 1/\sqrt{|T_{\mu \nu}|} \sim 1/\sqrt{\cal E} \ .
\label{con:time}
\eeq
Equations~(\ref{cond:linear}), (\ref{cond:matter}) and (\ref{con:time}) are 
the conditions that must be satisfied for our solution to be 
valid.

We orient the axes so that the (plane) gravitational wave is propagating 
in the $z$ direction. In the transverse and traceless (TT) gauge, the metric 
perturbation $h_{\mu \nu}$ takes the form
\beq
  h_{\mu \nu}(t,z) = \left[ \begin{array}{cccc}
 0 & 0 & 0 & 0 \\
 0 & h_+(t,z) & h_{\times}(t,z) & 0 \\
 0 & h_{\times}(t,z) & -h_+(t,z) & 0 \\
 0 & 0 & 0 & 0 \\
 \end{array} \right] \ ,
\label{eq:hab}
\eeq
where $h_+$ and $h_{\times}$ are the plus- and cross-polarization of the 
gravitational wave, respectively.
The inverse of $g_{\mu \nu}$ is $g^{\mu \nu}=\eta^{\mu \nu} - h^{\mu \nu}$,
where $h^{\mu \nu} = \eta^{\mu \alpha} \eta^{\nu \beta} h_{\alpha \beta}$. 
The metric determinant is $g = -1 + O(|h_{\mu \nu}|^2)$.
In the presence of the gravitational wave, we write
\beq
  B^i = B_0^i + \delta B^i \ , \ \ \ \
P=P_0 + \delta P \ , \ \ \ \ v^i = \delta v^i \ ,
\eeq
where the subscript 0 denotes the unperturbed quantity, and 
where the perturbation in density $\delta \rho$ will be related 
to $\delta P$ below.

In order to expand the stress-energy tensor of Eq.~(\ref{Tab:ba}) to 
linear order, we write
\beqn
  \sqrt{4\pi} \, b^t &=& g_{\mu i} B^i u^{\mu} =
B_0^i \delta v_i \ , \label{eq:b0} \\
  \sqrt{4\pi} \, b^i &=&  \frac{B^i}{u^0} + \sqrt{4\pi}\, b^t v^i 
= B_0^i + \delta B^i \ , \\
  4\pi b^2 &=& B_0^2
+ 2 B_{0i}\delta B^i \cr
 & & + h_+ [ (B_0^x)^2 - (B_0^y)^2 ]  + 2 h_{\times} B_0^x B_0^y  \ , \\
  \sqrt{4\pi} \, b_x &=&  B_0^x + \delta B^x + h_+ B_0^x
+ h_{\times} B_0^y \ , \\
  \sqrt{4\pi} \, b_y &=& B_0^y + \delta B^y
- h_+ B_0^y + h_{\times} B_0^x \ , \\
  \sqrt{4\pi} \, b_z &=& B_0^z + \delta B^z \ , \label{eq:b_z}
\eeqn
where $B_0 = \sqrt{(B_0^x)^2 + (B_0^y)^2 + (B_0^z)^2}$.
The evolution of $h_{\mu \nu}$ is determined by the linearized Einstein
equation (in the TT gauge):
\beqn
  \Box h_+ &=& -8\pi (T^{xx}-T^{yy}) \cr
 &=& 4 (B_0^x \delta B^x - B_0^y \delta B^y) + 16\pi \left( P_0
 + \frac{B_0^2}{8\pi} \right) h_+ \ , \ \ \ \ \   \\
  \Box h_{\times} &=& -8\pi (T^{xy}+T^{yx}) \cr
 &=& 4 (B_0^x \delta B^y + B_0^y \delta B^x )  
 + 16\pi \left( P_0 + \frac{B_0^2}{8\pi} \right) h_{\times} \ , \ \ \  \ \ 
\eeqn
where we have used (\ref{Tab:ba}) and (\ref{eq:b0})--(\ref{eq:b_z}) 
and where $\Box$ denotes the flat-spacetime D'Alembertian operator. 
The left hand sides of the above equations are of order $k^2 h_0$ 
and hence much greater than the right hand sides by virtue of 
condition~(\ref{cond:matter}). Thus, we have
\beqn
  \Box h_+ &=& 0 \ , \label{lineq:hplus} \\
  \Box h_{\times} &=& 0 \ . \label{lineq:hcross}
\eeqn
To the order we are considering, the gravitational wave is unaffected
by the magnetized fluid. 

In this approximation, the constraint equation~(\ref{divB}) simplifies to
\beq
  \partial_i \delta B^i = 0 \ ,
\label{lineq:divB}
\eeq
and the linearized induction equation takes the form 
\beq
  \partial_t \delta B^i = -B_0^i \partial_j \delta v^j
  + B_0^j \partial_j \delta v^i \ . \label{lineq:induction}
\eeq
Since the gravitational wave is weak, the induced fluid motion does not 
develop shocks and thus the flow is adiabatic. Hence the solution 
to the energy equation is simply adiabatic compression, for which 
$P \propto \rho_0^{\Gamma}$. It follows that $\delta P = (\Gamma P/\rho_0) 
\delta \rho_0$ and the linearized continuity equation may be written as 
\beq
  \partial_t \delta P = -\Gamma P_0 \partial_j \delta v^j \ .
\label{lineq:continuity}
\eeq
The momentum equation~(\ref{eq:energy-momentum}), setting $\nu=i$, becomes
\beqn
  \partial_t \left( {\cal E} \delta v_i - \frac{B_{0i} B_{0j}}{4\pi} \delta v^j
\right) &=& -\partial_i \delta P - \frac{B_{0j}\partial_i \delta B^j}{4\pi} \cr
 & & + \frac{B_0^j\partial_j \delta B_i}{4\pi} + S^{GW}_i \ , \ \ 
\label{lineq:mom0}
\eeqn
where the gravitational wave source term is
\beqn
  & 4\pi S^{GW}_i = 
B_0^z (B_0^x \partial_z h_+ + B_0^y \partial_z h_{\times})
\delta_i{}^x  \ \ \ \ \  &  \cr \cr
  & + B_0^z (-B_0^y \partial_z h_+ + B_0^x \partial_z h_{\times})
\delta_i{}^y  & \cr \cr
  & + \{ [(B_0^y)^2 - (B_0^x)^2] \partial_z h_+
 - 2 B_0^x B_0^y \partial_z h_{\times}\} \delta_i{}^z \ , & 
\label{gw_source}
\eeqn
and where we have used Eq.~(\ref{lineq:divB}).
Note that if $B_0^x = B_0^y=0$, then $S^{GW}_i=0$. The gravitational
wave has no effect on the fluid in this case (in the TT gauge). 
In general, from the perturbation 
equations~(\ref{lineq:divB})--(\ref{lineq:mom0}), we see that to 
linear order all the three-vector indices can be raised and lowered 
by the flat-spacetime metric. Therefore, we hereafter regard 
all three-vectors in the perturbation equations as residing in 
the flat spacetime. It is easy to show that 
\beq
  \ve{S}^{GW} = \ve{j}_h \times \ve{B}_0 \ , 
\label{eq:veS}
\eeq
where 
\beqn
  j_h^x &=& \frac{1}{4\pi} (B_0^y \partial_z h_+ 
-B_0^x \partial_z h_{\times})  \ , \\
  j_h^y &=& \frac{1}{4\pi} (B_0^x \partial_z h_+ 
+ B_0^y \partial_z h_{\times}) \ , \\
  j_h^z &=& 0 \ .
\eeqn
Equations~(\ref{lineq:induction})--(\ref{gw_source}) with the 
constraint~(\ref{lineq:divB}) are the linearized MHD equations in 
the presence of the gravitational wave $h_{\mu \nu}$. 

Note that our equations are expressed in a coordinate basis. 
It is easy to transform the equations to other basis vectors. 
In Ref.~\cite{moortgat04}, the authors derived the MHD equations 
in the orthonormal tetrad basis using a different approach. We have 
verified that our equations~(\ref{lineq:induction}), (\ref{lineq:continuity}) 
and (\ref{lineq:mom0}) are equivalent to their equations~(14), (22) and 
(24), respectively [note that the authors assume $B_0^y=0$, and that 
a minus sign is missing in their equation (16) for $\ve{j}_E$]. 
Since all of the GR MHD codes to date evolve the coordinate components of 
the MHD variables, we will continue to work in the coordinate frame 
for the purpose of setting up a code-test problem.

\section{Analytic solution}
\label{sec:solution}

For simplicity, we assume that $h_+$ and $h_{\times}$ take the form 
of standing waves:
\beqn
   h_+(t,z) &=& h_{+0} \sin \! kz \cos \! kt \ , \label{eq:hplus} \\ 
  h_{\times}(t,z) &=& h_{\times 0} \sin \! kz \cos \! kt \ , 
\label{eq:hcross}
\eeqn
where $h_{+0}$ and $h_{\times 0}$ are constants. It is easy to 
verify that these forms of $h_+$ and $h_{\times}$ satisfy Eqs.~(\ref{lineq:hplus}) 
and~(\ref{lineq:hcross}). One may also show that a rotation by an 
angle $\varphi = \frac{1}{2}\tan^{-1} (h_{\times 0}/h_{+0})$ 
about the $z$-axis will make 
$h_{\times}(t,z)=0$ and $h_+(t,z)=h_0\,
\sin \! kz \cos \! kt$ in the new coordinate system, 
where $h_0=\sqrt{h_{+0}^2 +h_{\times 0}^2}$. 
Similarly, a rotation by an angle $\varphi = -\frac{1}{2}\tan^{-1} 
(h_{+0}/h_{\times 0})$ will make $h_+(t,z)=0$ and 
$h_{\times}(t,z)=h_0\, \sin \! kz \cos \! kt$
in the new coordinate 
system. This means that setting either $h_{+0}=0$ or $h_{\times 0}=0$ will not 
sacrifice generality for forms of $h_+$ and $h_{\times}$ given in 
Eqs.~(\ref{eq:hplus}) and (\ref{eq:hcross}).
However, keeping both $h_{+0}$ and $h_{\times 0}$ 
may still be useful for the purpose of testing a numerical code.

Substituting Eqs.~(\ref{eq:hplus}) and (\ref{eq:hcross}) into 
Eq.~(\ref{gw_source}), the gravitational-wave source term, $S^{GW}_i$, 
is given by
\beqn
  4\pi S^{GW}_x &=& k B_0^z (B_0^x h_{+0} + B_0^y h_{\times 0}) 
\cos \! kz \cos \! kt \ , \label{eq:sxgw} \\
  4\pi S^{GW}_y &=& k B_0^z (-B_0^y h_{+0} + B_0^x h_{\times 0}) 
\cos \! kz \cos \! kt \ , \label{eq:sygw} \\
  4\pi S^{GW}_z &=& -k \{ [ (B_0^x)^2 - (B_0^y)^2 ] h_{+0} + 2 
B_0^x B_0^y h_{\times 0} \} \times \cr 
& &  \cos \! kz \cos \! kt \ . \label{eq:szgw}
\eeqn
The time derivative of $S^{GW}_i$ can be written as 
\beqn
  \partial_t S^{GW}_i &=& -k^2 \cos \! kz \sin \! kt \, \tilde{S}^{GW}_i \ , \\
  4\pi \tilde{S}^{GW}_x &=& (B_0^x h_{+0} + B_0^y h_{\times 0})
B_0^z \ , \label{eq:stxgw} \\ 
  4\pi \tilde{S}^{GW}_y &=& (B_0^x h_{\times 0} - B_0^y h_{+0})
B_0^z \ , \label{stygw} \\
  4\pi \tilde{S}^{GW}_z &=& -[(B_0^x)^2 - (B_0^y)^2] h_{+0} 
- 2 B_0^x B_0^y h_{\times 0} \ . \ \  \label{eq:stzgw}
\eeqn
The solution of the linearized 
MHD equations is uniquely determined by the initial 
conditions. We consider the case where all the MHD variables take 
their unperturbed values at $t=0$ and hence
\beq
  \delta B^i(0,\ve{x}) = 0 \ \ , \ \ 
\delta v^i(0,\ve{x}) = 0 \ \ , \ \ 
\delta P(0,\ve{x}) =0 \ .
\label{cond:initial}
\eeq
It follows from Eqs.~(\ref{lineq:mom0}) and~(\ref{cond:initial}) that 
\beq
  \left. \partial_t [ \delta \ve{v} - \ve{v}_A (\ve{v}_A \cdot \delta \ve{v}) 
 ] \right|_{t=0} = \frac{\ve{S}_0^{GW}}{\cal E} \ ,
\label{eq:nameless}
\eeq
where we have introduced the Alfv\'en velocity $\ve{v}_A$, 
\beq
  \ve{v}_A = \frac{\ve{B}_0}{\sqrt{4\pi \cal E}} \ , 
\eeq
and where 
\beq
  \ve{S}_0^{GW} = \ve{S}^{GW}(t=0)=k \cos \! kz\, \tilde{\ve{S}}^{GW} \ .
\label{eq:SGW0}
\eeq
Contracting the above equation with $\ve{v}_A$ gives 
\beq
  \left. \partial_t (\ve{v}_A \cdot \delta \ve{v}) 
\right|_{t=0} = 0 \ , \label{eq:vadotvt0}
\eeq
where we have used the fact that $\ve{v}_A \cdot \ve{S}^{GW}=0$ by virtue 
of Eq.~(\ref{eq:veS}).
Substituting Eq.~(\ref{eq:vadotvt0}) to Eq.~(\ref{eq:nameless}) gives 
\beq
  \left. \partial_t \delta \ve{v} \right|_{t=0} = 
\frac{\ve{S}_0^{GW}}{\cal E} \ .
\label{eq:dtdvt0}
\eeq

Taking the time derivative of Eq.~(\ref{lineq:mom0}) and using 
Eqs.~(\ref{lineq:induction}) and~(\ref{lineq:continuity}), we obtain
\beqn
 \partial_t^2 \left( {\cal E} \delta v_i - \frac{B_{0i} B_{0j}}{4\pi}
\delta v^j\right)
 = \left( \Gamma P_0 + \frac{B_0^2}{4\pi} \right) \partial_i \partial_j 
\delta v^j & &   \cr \cr
 - \frac{B_{0j} B_0^k}{4\pi} \partial_i \partial_k \delta v^j 
  - \frac{B_{0i}B_0^j}{4\pi} \partial_j \partial_k \delta v^k  \cr \cr
  + \frac{B_0^j B_0^k}{4\pi} \partial_j \partial_k \delta v_i 
   + \partial_t S_i^{GW} \ . \ \ \ \ \ & &
\label{lineq:mom}
\eeqn

We first find a particular solution that solves Eq.~(\ref{lineq:mom}) 
by writing 
\beq
\delta \ve{v}(t,z) = \delta \tilde{\ve{v}}_p 
\cos \! kz \sin \! kt \ . 
\label{part-sol}
\eeq
Equation~(\ref{lineq:mom}) becomes 
\beqn
  [ k^2 - (\ve{k} \cdot \ve{v}_A)^2] \delta \tilde{\ve{v}}_p 
+ [ (\ve{k} \cdot \ve{v}_A) (\ve{k} \cdot \delta \tilde{\ve{v}}_p) 
-k^2 (\ve{v}_A \cdot \delta \tilde{\ve{v}}_p) ] \ve{v}_A  & & \cr \cr
+ [ (\ve{k} \cdot \ve{v}_A) (\ve{v}_A \cdot \delta \tilde{\ve{v}}_p) 
-c_m^2 (\ve{k} \cdot \delta \tilde{\ve{v}}_p) ] \ve{k} = \frac{k^2}{w}
\tilde{\ve{S}}^{GW} \ , \ \ \ \ \  &  & 
\label{lineq:momgw}
\eeqn 
where $\ve{k} = k \hat{\ve{z}}$. The quantity $c_m^2$ is defined as  
\beq
c_m^2 = \frac{\Gamma P_0 + B_0^2/4\pi}{\cal E} = v_A^2 + c_s^2 (1-v_A^2) \ ,
\label{def:cm}
\eeq
where $c_s = \sqrt{\Gamma P_0/(\rho_0 + P_0 + \rho_0 \epsilon_0)}$ is 
the sound speed. Contracting Eq.~(\ref{lineq:momgw}) with 
$\ve{k}$, we obtain 
\beq
  \delta \tilde{v}_p^z = \frac{\tilde{S}_{GW}^z}{{\cal E} (1-c_m^2)} \ .
\label{part:dvz}
\eeq
The other components of $\delta \tilde{\ve{v}}_p$ are determined by 
the $x$ and $y$ components of Eq.~(\ref{lineq:momgw}). The result is
\beqn
  \delta \tilde{v}_p^j &=& -\frac{v_A^z 
\tilde{S}^{GW}_z}{{\cal E}(1-v_A^2)[ 1-(v_A^z)^2]} v_A^j  \cr \cr
& & + \frac{\tilde{S}_{GW}^j}{{\cal E} [1-(v_A^z)^2]} \ \ \ , 
\ \ (j=x \, , \, y) \, .  \label{part:dvk}
\eeqn
Note that this particular solution satisfies the conditions 
$\delta \ve{v}(0,\ve{x}) = 0$, but it does not satisfy the 
conditions~(\ref{eq:dtdvt0}) required by the initial data~(\ref{cond:initial}). 
Therefore, we need to add a suitable homogeneous solution (i.e., a solution 
of Eq.~(\ref{lineq:mom}) with $S^{GW}_i=0$) to the particular solution 
to comply with conditions~(\ref{eq:dtdvt0}).

To find the homogeneous solutions, we decompose $\delta \ve{v}$ into 
Fourier modes,
\beq
  \delta \ve{v}(t,\ve{x}) = \int \delta \tilde{\ve{v}}(\ve{\kappa}) 
e^{i (\ve{\kappa} \cdot \ve{x} - \omega t)} d^3 \kappa 
\label{eq:dvfourier}
\eeq
and substitute Eq.~(\ref{eq:dvfourier}) into Eq.~(\ref{lineq:mom}), 
with $S^{GW}_i=0$. We obtain 
\beqn
  [ \omega^2 - (\ve{\kappa} \cdot \ve{v}_A)^2] \delta \tilde{\ve{v}} + 
[ (\ve{\kappa} \cdot \ve{v}_A) ( \ve{\kappa} \cdot \delta \tilde{\ve{v}})  
- \omega^2 (\ve{v}_A \cdot \delta \tilde{\ve{v}}) ] \ve{v}_A & & \cr 
+ [ (\ve{\kappa} \cdot \ve{v}_A) (\ve{v}_A \cdot \delta \tilde{\ve{v}}) 
- c_m^2 (\ve{\kappa} \cdot \delta \tilde{\ve{v}})] 
\ve{\kappa} = 0 \ . \ \ \ \ \ & & 
\label{lineq:momw}
\eeqn 
The homogeneous solutions are the nontrivial 
solutions of Eq.~(\ref{lineq:momw}), which are well known. The solutions 
consist of three modes corresponding to an Alfv\'en wave and 
both fast and slow magnetosonic waves (the entropy wave is absent 
for isentropic flows). The eigenvalues and eigenvectors when 
$\ve{\kappa} \cdot \ve{v}_A \neq 0$ are as follows
(see, e.g.~\cite{komissarov99}):

{\it Alfv\'en wave (with $\ve{\kappa} \cdot \ve{v}_A \neq 0$)}: 
\beqn
  \omega^2 = \omega_A^2 &\equiv & 
(\ve{\kappa}\cdot \ve{v}_A)^2 \ ,  \\
 \delta \tilde{\ve{v}}(\ve{\kappa}) \propto \tilde{\ve{u}}_A(\ve{\kappa}) &\equiv & 
\ve{\kappa} \times \ve{v}_A \ ; \label{eigen:alfven}
\eeqn

{\it Fast and slow magnetosonic waves 
(with $\ve{\kappa} \cdot \ve{v}_A \neq 0$)}: 
\beqn
  \omega^2 = \omega_m^2 \ , \ 
\mbox{where $\omega_m^2$ are the roots of the dispersion equation} & &  \cr \cr
  \omega_m^4 - [ \kappa^2 c_m^2 + c_s^2 (\ve{\kappa} \cdot 
\ve{v}_A)^2 ] \omega_m^2
 + \kappa^2 c_s^2 (\ve{\kappa} \cdot \ve{v}_A)^2 = 0 \ , 
\ \ \ \ \ & & \label{eq:dis} \\ \cr
 \mbox{[The fast (slow) mode is the one with larger (smaller)}  & &  \cr 
 \mbox{value of $\omega_m^2$.]} \ \ \ \ \ \ \ \ \ \ \ \ \ \ \ \ 
\ \ \ \ \ \ \ \ \ \ \ \ \ \ \ \ \ \ \ \ \ \ \ \ \ \ \ \ \ \ \ \ 
\ \ \ \  & & \cr \cr
 \delta \tilde{\ve{v}}(\ve{\kappa}) \propto \tilde{\ve{u}}_m(\ve{\kappa}) 
\equiv \ve{v}_A 
+ \frac{\omega_m^2 (1-v_A^2)}{(\omega_m^2-\kappa^2)(\ve{\kappa} \cdot \ve{v}_A)}
 \ve{\kappa} \ . \ \ \ \ \ \ \ \ \ \  & & 
\label{eigen:magneto}  
\eeqn

When $\ve{\kappa} \cdot \ve{v}_A=0$, the Alfv\'en mode and one of the 
magnetosonic modes become two linearly independent static modes 
($\omega^2=0$). In this case, the 
three independent modes are given as follows: 

{\it Magnetosonic wave (with $\ve{\kappa} \cdot \ve{v}_A = 0$)}:
\beqn
  \omega^2 &=& c_m^2 \kappa^2 \ , 
\label{eq:magnetosonicw0} \\ 
  \delta \tilde{\ve{v}}(\ve{\kappa}) & \propto & \ve{\kappa} \ ;
\label{eq:magnetosonicv0}
\eeqn

{\it Two static modes ($\ve{\kappa} \cdot \ve{v}_A = 0$)}:
\beqn
  \omega^2 &=& 0 \ , \\
  \delta \tilde{\ve{v}}(\ve{\kappa})  \propto  \ve{v}_A \ \ \ \ & , & \ \ \ \ 
  \delta \tilde{\ve{v}}(\ve{\kappa}) \propto \ve{\kappa} \times \ve{v}_A \ .
\eeqn

The general solution of Eq.~(\ref{lineq:mom}) can be written as 
\beqn
  \delta \ve{v}(t,\ve{x}) &=& \delta \ve{v}_p(t,z) + 
\int d^3 \kappa \mathcal{A}_A(\ve{\kappa}) \tilde{\ve{u}}_A(\ve{\kappa}) 
e^{i(\ve{\kappa} \cdot \ve{x}-\omega_A t)} \cr
  & & + \sum_{j=1}^2 \int d^3 \kappa \mathcal{A}_{mj}(\ve{\kappa}) 
  \tilde{\ve{u}}_{mj}(\ve{\kappa}) 
e^{i(\ve{\kappa} \cdot \ve{x}-\omega_{mj} t)} \\ 
  &=& \delta \ve{v}_p(t,z) + \int d^3 \kappa \, \tilde{\ve{u}}_A(\ve{\kappa}) 
[ C_A(\ve{\kappa}) 
\cos (\ve{\kappa} \cdot \ve{x}-\omega_A t)   \cr
  & &  + D_A(\ve{\kappa}) \sin (\ve{\kappa} \cdot \ve{x}-\omega_A t) ]  \cr 
 & & + \sum_{j=1}^2 \int d^3 \kappa \, \tilde{\ve{u}}_{mj}(\ve{\kappa})
[C_{mj}(\ve{\kappa}) \cos (\ve{\kappa} \cdot \ve{x}-\omega_{mj}t) \cr 
 & &  + D_{mj}(\ve{\kappa}) \sin (\ve{\kappa} \cdot \ve{x}-\omega_{mj} t) ] 
\ , \label{sol:general}
\eeqn
where $\delta \ve{v}_p(t,z)=\delta \ve{\tilde{v}}_p \cos kz \sin kt$,
$\omega_A = |\ve{\kappa} \cdot \ve{v}_A|$, $\tilde{\ve{u}}_A(\ve{\kappa})
=\ve{\kappa} \times \ve{v}_A$, $\omega_{mj}$ ($j=$1, 2)
are the two solutions of Eq.~(\ref{eq:dis}). The vectors 
$\tilde{\ve{u}}_{mj}(\ve{\kappa})$
are given by Eq.~(\ref{eigen:magneto}) with $\omega_m$ set to $\omega_{mj}$.
The complex functions $\mathcal{A}_A(\ve{\kappa})$ and 
$\mathcal{A}_{mj}(\ve{\kappa})$ are to be 
determined by the initial conditions. The functions $C_A(\ve{\kappa})$, 
$D_A(\ve{\kappa})$, $C_{mj}(\ve{\kappa})$ and $D_{mj}(\ve{\kappa})$ 
are linear combinations of $\mathcal{A}_A(\ve{\kappa})$ 
and $\mathcal{A}_{mj}(\ve{\kappa})$. We note that 
$\tilde{\ve{u}}_A(-\ve{\kappa})= 
-\tilde{\ve{u}}_A(\ve{\kappa})$ and $\tilde{\ve{u}}_{mj}(-\ve{\kappa})
=\tilde{\ve{u}}_{mj}(\ve{\kappa})$. Hence the stationary wave solutions 
are obtained by setting $C_A(-\ve{\kappa})=\pm C_A(\ve{\kappa})$, 
$D_A(-\ve{\kappa})=\pm D_A(\ve{\kappa})$, 
$C_{mj}(-\ve{\kappa})=\pm C_{mj}(\ve{\kappa})$, 
and $D_{mj}(-\ve{\kappa})=\pm D_{mj}(\ve{\kappa})$.

Since both $\delta \ve{v}_p$ and 
$\ve{S}_0^{(GW)}$ have $\cos \! kz$ spatial dependence, 
we need to add to $\delta \ve{v}_p$ a homogeneous solution with 
$\ve{\kappa} = \pm \ve{k}$ in order to satisfy the initial conditions 
[$\delta \ve{v}(0,\ve{x})=0$ and Eq.~(\ref{eq:dtdvt0})]. Specifically, 
if $\ve{k}\cdot \ve{v}_A \neq 0$ (we will consider the case 
$\ve{k}\cdot \ve{v}_A=0$ in the next subsection), we set 
\beqn
  C_A(\ve{\kappa}) &=& \frac{A_A}{2} [ \delta(\ve{\kappa}-\ve{k}) + 
\delta(\ve{\kappa}+\ve{k}) ] \ , \\ 
  D_A(\ve{\kappa}) &=& 0 \ , \\
  C_{mj}(\ve{\kappa}) &=& \frac{A_{mj}}{2} [ \delta(\ve{\kappa}-\ve{k}) 
-\delta(\ve{\kappa}+\ve{k}) ] \ , \\
  D_{mj}(\ve{\kappa}) &=& 0 \ ,
\eeqn
where $A_A$ and $A_{mj}$ are constants to be determined. 
Equation~(\ref{sol:general}) becomes 
\beqn
  \delta \ve{v}(t,z) &=& \delta \tilde{\ve{v}}_p \cos \! kz \sin \! kt
+ A_A \tilde{\ve{u}}_A(\ve{k}) \cos \! kz \sin \! \omega_A t \cr
 & & + A_{m1} \tilde{\ve{u}}_{m1}(\ve{k}) \cos \! kz
\sin \! \omega_{m1} t \cr 
 & & + A_{m2} \tilde{\ve{u}}_{m2}(\ve{k}) \cos \! kz
\sin \! \omega_{m2} t \ .
\eeqn
We see that this solution satisfies the initial conditions 
$\delta \ve{v}(0,\ve{x})=0$. The values of $A_A$, $A_{m1}$ and 
$A_{m2}$ are uniquely determined by Eq.~(\ref{eq:dtdvt0}). 
Because of the nature of the eigenmodes, we 
consider two separate cases in the following two subsections: 
$\ve{k} \cdot \ve{v}_A=0$ (i.e., $B_0^z =0$) and $\ve{k} \cdot \ve{v}_A \neq 0$ 
(i.e., $B_0^z \neq 0$).

\subsection{Case 1: $B_0^z =0$ ($\ve{k} \cdot \ve{v}_A=0$)}

In this case, the unperturbed magnetic field is perpendicular to 
the $\ve{k}$ vector associated with the standing wave. It follows from 
Eqs.~(\ref{eq:sxgw})--(\ref{eq:szgw}), (\ref{eq:SGW0}), (\ref{cond:initial}) 
and~(\ref{eq:dtdvt0}) that 
\beqn
  \delta v^x(t,\ve{x}) &=& \delta v^y(t,\ve{x})=0 \ ,
\label{eq:dvxdvy} \\ 
  \left. \partial_t \delta v^z(t,\ve{x}) \right|_{t=0} &=& 
  \frac{k \tilde{S}_{GW}^z}{\cal E}\cos \! kz \ .
\label{eq:dtdvzt0}
\eeqn
The particular solution is given by [see Eqs.~(\ref{part-sol}), 
(\ref{part:dvz}) and~(\ref{part:dvk})] 
\beq
  \delta \ve{v}_p = \hat{\ve{z}} \frac{\tilde{S}^z_{GW}}{{\cal E}(1-c_m^2)} 
\cos \! kz \sin \! kt \ .
\eeq
Hence, the solution satisfying Eqs.~(\ref{eq:dvxdvy}) and~(\ref{eq:dtdvzt0}) 
is obtained by adding a magnetosonic wave 
[Eqs.~(\ref{eq:magnetosonicw0}) 
and~(\ref{eq:magnetosonicv0})] to $\delta \ve{v}_p$: 
\beq
  \delta \ve{v}(t,\ve{x}) = \hat{\ve{z}} \left[ 
A_m \sin \! k c_m t + \frac{\tilde{S}^z_{GW}}{{\cal E}(1-c_m^2)} 
\sin \! kt \right] \cos \! kz \ .
\label{case1:sol0}
\eeq
The constant $A_m$ is determined by the remaining initial 
conditions~(\ref{eq:dtdvzt0}). 
The final solution is 
\beq
  \delta \ve{v} (t,\ve{x}) = -\hat{\ve{z}}\frac{\tilde{S}_{GW}^z}
{{\cal E}(1-c_m^2)} (c_m \sin \! kc_m t - \sin \! kt) \cos kz \ .
\label{case1:solv}
\eeq
The perturbations $\delta P$ and $\delta \ve{B}$ can be calculated 
by integrating Eqs.~(\ref{lineq:induction}) and~(\ref{lineq:continuity}). 
The result is 
\beqn
  \delta P(t,\ve{x}) &=& \frac{\Gamma P_0 \tilde{S}^{GW}_z}
{{\cal E}(1-c_m^2)} 
(\cos \! kc_m t - \cos \! kt) \sin \! kz \ , \label{case1:solP} \\ 
  \delta \ve{B}(t,\ve{x}) &=& \frac{\tilde{S}^{GW}_z \ve{B}_0}
{{\cal E}(1-c_m^2)} 
(\cos \! kc_m t - \cos \! kt) \sin \! kz \ ,  \label{case1:solB} \\ 
  &=& [\delta B_m(t,z)+\delta B_p(t,z) ] \hat{\ve{B}}_0 \ ,
\eeqn
where $\hat{\ve{B}}_0 = \ve{B}_0/B_0$ and 
\beqn
  \delta B_m(t,z) &=& \frac{\tilde{S}^{GW}_z B_0}{{\cal E}(1-c_m^2)} 
\cos \! kc_m t \sin \! kz \ , \\ 
  \delta B_p(t,z) &=& -\frac{\tilde{S}^{GW}_z B_0}{{\cal E}(1-c_m^2)} 
\cos \! kt \sin \! kz \ .
\eeqn

\begin{figure}
\includegraphics[width=5cm,angle=270]{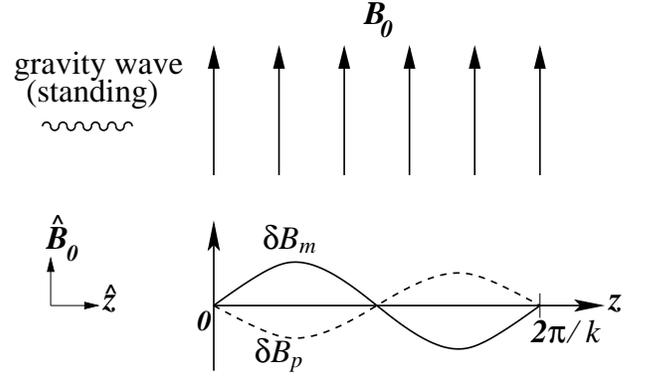}
\caption{Perturbation of the magnetic field $\delta \ve{B}=(\delta B_m 
+ \delta B_p) \hat{\ve{B}}_0$ induced by a standing gravitational wave when 
$\ve{k} \cdot \ve{v}_A=0$. The 
unperturbed magnetic field $\ve{B}_0$ is perpendicular to the 
wave vector of the gravitational wave $\ve{k}=k\hat{\ve{z}}$. Both stationary
modes $\delta B_p$ and $\delta B_m$ have the same $\sin \!kz$ spatial 
dependence, but they oscillate with different 
amplitudes and at different frequencies.}
\label{fig:Bpend}
\end{figure}

We see that with our initial data~(\ref{cond:initial}), 
the gravitational wave only induces a magnetosonic wave in 
this case, but not the other two static modes. 
Figure~\ref{fig:Bpend} shows schematically the perturbation 
$\delta \ve{B} = (\delta B_m + \delta B_p) \hat{\ve{B}}_0$. 
Both stationary modes $\delta B_m$ and 
$\delta B_p$ have $\sin \! kz$ spatial dependence, but they 
have different amplitudes and oscillate at different frequencies.

As an interesting case, 
we note from Eq.~(\ref{eq:stzgw}) that the gravitational wave 
will have no effect (in the adopted TT gauge) on the fluid or 
magnetic field if $\tilde{S}^z_{GW}=0$, i.e.\ if
$B_0^x$ and $B^y_0$ satisfies the equation 
\beq
  [(B_0^y)^2 - (B_0^x)^2] h_{+0} - 2B_0^x B_0^y h_{\times 0} = 0 \ .
\label{cond:SGWz0}
\eeq

\subsection{Case 2: $B_0^z \neq 0$ ($\ve{k}\cdot \ve{v}_A \neq 0$)}

In this case, we expect the gravitational wave will induce all the three 
modes. As discussed earlier, the solution that satisfies 
$\delta \ve{v}(0,\ve{x})=0$ can be written as 
\beqn
  \delta \ve{v}(t,z) &=& \delta \tilde{\ve{v}}_p \cos \! kz \sin \! kt
 + A_A \tilde{\ve{u}}_A \cos \! kz 
\sin \! \omega_A t \cr 
& & + A_{m1} \tilde{\ve{u}}_{m1} \cos \! kz
\sin \! \omega_{m1} t \cr
 & & + A_{m2} \tilde{\ve{u}}_{m2} \cos \! kz
\sin \! \omega_{m2} t \ .
\label{case2:solv}
\eeqn
To satisfy condition~(\ref{eq:dtdvt0}), we must choose the amplitudes 
$A_A$, $A_{m1}$ and $A_{m2}$ appropriately. We note from 
the definition of $\tilde{\ve{u}}_A$, $\tilde{\ve{u}}_{m1}$ 
and $\tilde{\ve{u}}_{m2}$ that $\tilde{\ve{u}}_A \cdot \tilde{\ve{u}}_{m1} 
= \tilde{\ve{u}}_A \cdot \tilde{\ve{u}}_{m2}=0$ and 
\beqn
  \left. \partial_t (\tilde{\ve{u}}_A \cdot \delta \ve{v}) \right|_{t=0} 
&=& [A_A |\tilde{\ve{u}}_A|^2 \omega_A + k (\tilde{\ve{u}}_A \cdot 
\delta \tilde{\ve{v}}_p) ] \cos \! kz \ , \ \ \ \ \ \label{eq:uadotdv}  \\
  \left. \partial_t (\ve{v}_A \cdot \delta \ve{v}) \right|_{t=0} 
&=& [ A_{m1} \omega_{m1} (\ve{v}_A \cdot \tilde{\ve{u}}_{m1}) \cr 
 & & + A_{m2} \omega_{m2} (\ve{v}_A \cdot \tilde{\ve{u}}_{m2}) \cr
 & & + k (\ve{v}_A \cdot \delta \tilde{\ve{v}}_p)] \cos \! kz \ , \\
  \left. \partial_t (\ve{k} \cdot \delta \ve{v}) \right|_{t=0} 
&=& [ A_{m1} \omega_{m1} (\ve{k} \cdot \tilde{\ve{u}}_{m1}) \cr
& & + A_{m2} \omega_{m2} (\ve{k} \cdot \tilde{\ve{u}}_{m2}) \cr
& &  + k (\ve{k} \cdot \delta \tilde{\ve{v}}_p)] \cos \! kz \ .
\label{eq:kdotdv}
\eeqn
Define 
\beqn
  \xi_A &=& \frac{k}{\cal E} \tilde{\ve{u}}_A \cdot \tilde{\ve{S}}^{GW} 
\ , \label{eq:xi_A} \\
  \xi_k &=& \frac{k}{\cal E} \ve{k} \cdot \tilde{\ve{S}}^{GW} \ .
\label{eq:xi_k}
\eeqn
From Eqs.~(\ref{eq:SGW0})--(\ref{eq:dtdvt0}), we have
\beqn
 \left. \partial_t (\tilde{\ve{u}}_A \cdot \delta \ve{v}) \right|_{t=0}
&=& \xi_A \cos \! kz \ , \label{eq:uadotdv2}  \\
  \left. \partial_t (\ve{v}_A \cdot \delta \ve{v}) \right|_{t=0} 
&=& 0 \ ,  \\
 \left. \partial_t (\ve{k} \cdot \delta \ve{v}) \right|_{t=0} 
&=& \xi_k \cos \! kz \ . \label{eq:kdotdv2}
\eeqn
Hence, matching Eqs.~(\ref{eq:uadotdv})--(\ref{eq:kdotdv}) with 
(\ref{eq:uadotdv2})--(\ref{eq:kdotdv2}) yields
\beq
  A_A = \frac{\xi_A - k(\tilde{\ve{u}}_A \cdot 
\delta \tilde{\ve{v}}_p)}{|\tilde{\ve{u}}_A|^2 \omega_A} \ ,
\label{eq:AA}
\eeq
\beqn
   A_{m1} \omega_{m1} (\ve{v}_A \cdot \tilde{\ve{u}}_{m1})
+ A_{m2} \omega_{m2} (\ve{v}_A \cdot \tilde{\ve{u}}_{m2}) & & \cr
+ k (\ve{v}_A \cdot \delta \tilde{\ve{v}}_p) = 0 \ , & &
\label{eq:Am1} \\
A_{m1} \omega_{m1} (\ve{k} \cdot \tilde{\ve{u}}_{m1})
+ A_{m2} \omega_{m2} (\ve{k} \cdot \tilde{\ve{u}}_{m2}) & & \cr
 + k (\ve{k} \cdot \delta \tilde{\ve{v}}_p) = \xi_k \ . & & 
\label{eq:Am2}
\eeqn
The last two equations determine the values of $A_{m1}$ and $A_{m2}$. 
Specifically, 
\beqn
  A_{m1} &=& \frac{c_1 a_{22}- c_2 a_{12}}{a_{11} a_{22}-a_{12}a_{21}} 
\ , \label{eq:exAm1} \\
  A_{m2} &=& \frac{c_2 a_{11} - c_1 a_{21}}{a_{11} a_{22}-a_{12}a_{21}}
\label{eq:exAm2}
\eeqn
where
\beqn
  a_{11} &=& \omega_{m1} (\ve{v}_A \cdot \tilde{\ve{u}}_{m1}) \ , \\
  a_{12} &=& \omega_{m2} (\ve{v}_A \cdot \tilde{\ve{u}}_{m2}) \ , \\
  a_{21} &=& \omega_{m1} (\ve{k} \cdot \tilde{\ve{u}}_{m1}) \ , \\
  a_{22} &=& \omega_{m2} (\ve{k} \cdot \tilde{\ve{u}}_{m2}) \ , \\
  c_1 &=& -k (\ve{v}_A \cdot \delta \tilde{\ve{v}}_p) \ , \\
  c_2 &=& \xi_k- k (\ve{k} \cdot \delta \tilde{\ve{v}}_p) \ . 
\label{eq:c_2}
\eeqn

The perturbations $\delta P$ and $\delta B^i$ are easily obtained by 
integrating Eqs.~(\ref{lineq:continuity}) and~(\ref{lineq:induction}). 
The result is 
\beqn
  \delta P(t,z) &=& k \Gamma P_0 \left[ 
\frac{A_{m1}}{\omega_{m1}} 
\tilde{u}^z_{m1} (1-\cos \! \omega_{m1} t) \right.  \cr
 & & + \frac{A_{m2}}{\omega_{m2}} \tilde{u}^z_{m2} 
(1-\cos \! \omega_{m2} t)  \cr \cr
& & \left. + \frac{\delta \tilde{v}^z_p}{k} 
(1-\cos \! kt) \right] \sin \! kz \label{case2:solP} \\ 
  \delta \ve{B}(t,z) &=& k \left[ -\frac{B_0^z A_A}{\omega_A} 
\tilde{\ve{u}}_A (1-\cos \! \omega_A t) \right. \cr  \cr 
& & + \frac{A_{m1}}{\omega_{m1}} ( \tilde{u}^z_{m1} \ve{B}_0 
- B_0^z \tilde{\ve{u}}_{m1} ) (1-\cos \! \omega_{m1} t)  \cr \cr 
& & + \frac{A_{m2}}{\omega_{m2}} ( \tilde{u}^z_{m2} \ve{B}_0
- B_0^z \tilde{\ve{u}}_{m2} ) (1-\cos \! \omega_{m2} t)  \cr \cr
& & \left. + \frac{1}{k} (\delta \tilde{v}^z_p \ve{B}_0 - B_0^z \delta 
\tilde{\ve{v}}_p) (1-\cos \! kt) \right] \sin \! kz \cr  & & 
\label{case2:solB}
\eeqn

It can be shown that in the limit $B_0^z \rightarrow 0$, the 
solutions~(\ref{case2:solv}), (\ref{case2:solP}) and~(\ref{case2:solB}) 
reduce to Eqs.~(\ref{case1:solv}), (\ref{case1:solP}) and~(\ref{case1:solB}), 
respectively. It is also easy to show that if 
$\ve{k} \cdot \tilde{\ve{S}}^{GW}=0$ [which, from Eq.~(\ref{eq:szgw}), 
is the same condition as 
Eq.~(\ref{cond:SGWz0})], only the Alfv\'en wave is excited; if 
$(\ve{k} \times \ve{v}_A) \cdot \tilde{\ve{S}}^{GW}=0$, only the 
fast and slow magnetosonic waves can be excited.

\begin{figure}
\includegraphics[width=4.5cm,angle=270]{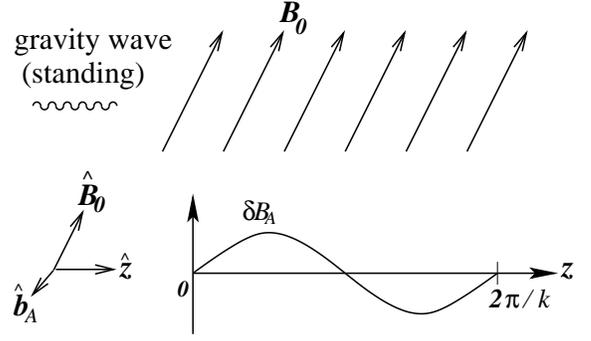}
\caption{Gravitation-wave induced Alfv\'en mode $\delta B_A \hat{\ve{b}}_A$ 
when $\ve{k} \cdot \ve{v}_A \neq 0$.
The direction of perturbation $\hat{\ve{b}}_A$ is perpendicular to both 
$\hat{\ve{B}}_0$ and $\hat{\ve{z}}$.} 
\label{fig:alfven}
\end{figure}

\begin{figure}
\includegraphics[width=6cm,angle=270]{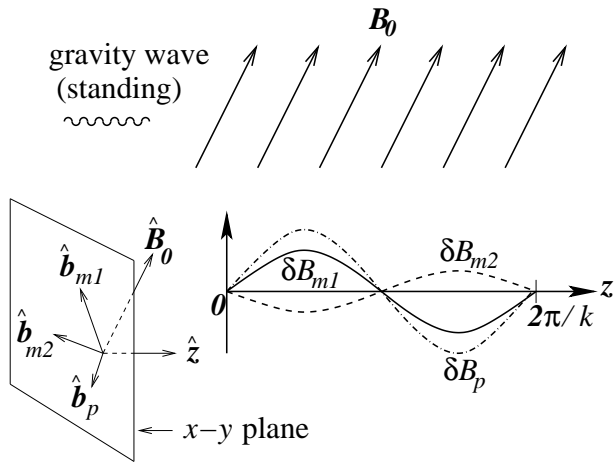}
\caption{Gravitation-wave induced magnetosonic modes 
($\delta B_{m1} \hat{\ve{b}}_{m1}$ and $\delta B_{m2} \hat{\ve{b}}_{m2}$) 
and the particular solution ($\delta B_p \hat{\ve{b}}_p$) 
when $\ve{k} \cdot \ve{v}_A \neq 0$. All three 
unit vectors $\hat{\ve{b}}_{m1}$, $\hat{\ve{b}}_{m2}$ and $\hat{\ve{b}}_p$ 
are perpendicular to $\hat{\ve{z}}$. All the standing modes have $\sin \!kz$ 
spatial dependence, but they oscillate with different amplitudes and at 
different frequencies.}
\label{fig:others}
\end{figure}

To visualize the perturbation of the magnetic field, we write 
\beq
  \delta \ve{B} = \delta B_A \hat{\ve{b}}_A + \delta B_{m1} 
\hat{\ve{b}}_{m1} + \delta B_{m2} \hat{\ve{b}}_{m2} 
+ \delta B_p \hat{\ve{b}}_p \ ,
\eeq
where $\hat{\ve{b}}_A$, $\hat{\ve{b}}_{m1}$, $\hat{\ve{b}}_{m2}$, 
and $\hat{\ve{b}}_p$ are unit vectors in the directions of 
$\tilde{\ve{u}}_A$, $\tilde{u}^z_{m1} \ve{B}_0- B_0^z \tilde{\ve{u}}_{m1}$, 
$\tilde{u}^z_{m2} \ve{B}_0 - B_0^z \tilde{\ve{u}}_{m2}$, and 
$\delta \tilde{v}^z_p \ve{B}_0 - B_0^z \delta \tilde{\ve{v}}_p$, 
respectively. All four unit vectors are perpendicular to $\hat{\ve{z}}$.
The expressions for $\delta B_A$, $\delta B_{m1}$, 
$\delta B_{m2}$ and $\delta B_p$ can be deduced from Eq.~(\ref{case2:solB}). 
Figure~\ref{fig:alfven} shows the Alfv\'en part of the perturbation 
($\delta B_A \hat{\ve{b}}_A$) and Fig.~\ref{fig:others} shows 
the other pieces of $\delta \ve{B}$. All of these four independent 
standing modes 
have the $\sin \! kz$ spatial dependence, but they have different 
amplitudes and oscillate with different frequencies.

Note that for $\ve{B}_0=0$ there is no pressure, density or 
velocity perturbation as measured in the coordinate (TT) gauge 
adopted here. Transforming to a proper orthonormal reference 
frame, where coordinate separations measure proper distances, 
the pressure and density perturbations remain zero (they are 
scalar invariants) but the fluid acquires a familiar 
quadrupolar velocity pattern orthogonal to $\ve{k}$~\cite{fn1}. 
Such a pattern conserves the comoving volume of a fluid element, 
and hence the density and pressure. 

\section{Summary}
\label{sec:summary}

\begin{table*}
\caption{Summary: Analytic solution}
\fbox{ \parbox{15cm}{
\mbox{\bf Gravitational wave:}
\begin{eqnarray*}
   h_+(t,z) &=& h_{+0} \sin \! kz \cos \! kt  \label{eq:sum-hplus} \\
  h_{\times}(t,z) &=& h_{\times 0} \sin \! kz \cos \! kt 
\label{eq:sum-hcross}
\end{eqnarray*}

\mbox{\bf MHD wave:}
\vskip 0.2cm
\mbox{Case~1: $B_0^z=0$}
\begin{eqnarray*}
  \delta P(t,z) &=& \frac{\Gamma P_0 \tilde{S}^{GW}_z}
{{\cal E}(1-c_m^2)}
(\cos \! kc_m t - \cos \! kt) \sin \! kz = 
\frac{\Gamma P_0}{\rho_0} \delta \rho_0  \\
  \delta \ve{B}(t,z) &=& \frac{\tilde{S}^{GW}_z \ve{B}_0}
{{\cal E}(1-c_m^2)}
(\cos \! kc_m t - \cos \! kt) \sin \! kz   \label{case1:sum-solB} \\
  \delta \ve{v} (t,z) &=& -\hat{\ve{z}}\frac{\tilde{S}_{GW}^z}
{{\cal E}(1-c_m^2)} (c_m \sin \! kc_m t - \sin \! kt) \cos kz 
\label{case1:sum-solv}
\end{eqnarray*}

Here $\tilde{S}^{GW}_z$, $\cal E$ and $c_m$ are given by 
Eqs.~(\ref{eq:stzgw}), (\ref{def:w}) and~(\ref{def:cm}), 
respectively.
\vskip 0.5cm
\mbox{Case~2: $B_0^z \neq 0$}
\begin{eqnarray*}
  \delta P(t,z) &=& k \Gamma P_0 \left[
\frac{A_{m1}}{\omega_{m1}}
\tilde{u}^z_{m1} (1-\cos \! \omega_{m1} t) 
 + \frac{A_{m2}}{\omega_{m2}} \tilde{u}^z_{m2}
(1-\cos \! \omega_{m2} t) \right.  \cr \cr
& & \left. + \frac{\delta \tilde{v}^z_p}{k}
(1-\cos \! kt) \right] \sin \! kz = 
\frac{\Gamma P_0}{\rho_0} \delta \rho_0 \\
  \delta \ve{B}(t,z) &=& k \left[ -\frac{B_0^z A_A}{\omega_A}
\tilde{\ve{u}}_A (1-\cos \! \omega_A t) \right. \cr  \cr
& & + \frac{A_{m1}}{\omega_{m1}} ( \tilde{u}^z_{m1} \ve{B}_0
- B_0^z \tilde{\ve{u}}_{m1} ) (1-\cos \! \omega_{m1} t)  \cr \cr
& & + \frac{A_{m2}}{\omega_{m2}} ( \tilde{u}^z_{m2} \ve{B}_0
- B_0^z \tilde{\ve{u}}_{m2} ) (1-\cos \! \omega_{m2} t)  \cr \cr
& & \left. + \frac{1}{k} (\delta \tilde{v}^z_p \ve{B}_0 - B_0^z \delta
\tilde{\ve{v}}_p) (1-\cos \! kt) \right] \sin \! kz \cr  & &
\label{case2:sum-solB} \\
  \delta \ve{v}(t,z) &=& \delta \tilde{\ve{v}}_p \cos \! kz \sin \! kt
 + A_A \tilde{\ve{u}}_A \cos \! kz
\sin \! \omega_A t \cr
& & + A_{m1} \tilde{\ve{u}}_{m1} \cos \! kz
\sin \! \omega_{m1} t
 + A_{m2} \tilde{\ve{u}}_{m2} \cos \! kz
\sin \! \omega_{m2} t 
\label{case2:sum-solv}
\end{eqnarray*}
Here $\delta \tilde{\ve{v}}_p$ is given by Eqs.~(\ref{part:dvz}) and 
(\ref{part:dvk}); $\tilde{\ve{u}}_A$ is given by Eq.~(\ref{eigen:alfven}) 
with $\ve{\kappa}=\ve{k}=k\hat{\ve{z}}$; $\tilde{\ve{u}}_{m1}$ and 
$\tilde{\ve{u}}_{m2}$ 
are computed from Eqs.~(\ref{eq:dis}) and (\ref{eigen:magneto}) with 
$\ve{\kappa}=\ve{k}$; $A_A$, $A_{m1}$ and $A_{m2}$ are computed from 
Eqs.~(\ref{eq:AA}), (\ref{eq:exAm1})--(\ref{eq:c_2}), (\ref{eq:xi_A}), 
(\ref{eq:xi_k}), (\ref{eq:stxgw})--(\ref{eq:stzgw}).
} }
\label{tab:sol}
\end{table*}

We have derived and solved the linearized equations for
standing gravitational waves oscillating in an initially
homogeneous,  
magnetized fluid (see Table~\ref{tab:sol}). We express the 
equations and analytic solutions in a coordinate basis so that the 
results can be compared directly with numerical data from a GR MHD 
code. 

Our linearized equations are valid when (1) the amplitude of the 
gravitational wave is small ($h_0 \ll 1$), and 
(2) the magnitude of the stress-energy tensor of the 
unperturbed fluid is sufficiently small that the 
perturbation it induces in the background, nearly Minkowski 
spacetime remains small 
compared to $h_0$. Specifically, if the domain 
in the fluid is $\sim 1/k$, where $k$ is the wave number of the 
gravitational wave, the second condition can be written as 
${\cal E} \ll k^2 h_0$ [Eq.~(\ref{cond:matter})]. Our solution 
remains valid for a time $t \ll 1/\sqrt{\cal E}$, after 
which the homogeneous background undergoes collapse. 

We consider the metric perturbations of 
the form in Eqs.~(\ref{eq:hab}), (\ref{eq:hplus}) 
and~(\ref{eq:hcross}). The gravitational wave is a standing wave whose 
amplitude varies in the $z$-direction. It
can be regarded as the superposition of two equal-amplitude,  
infinite traveling waves moving in 
the $\pm z$-directions. At $t=0$, the fluid is assumed to be 
homogeneous, uniformly magnetized, and at rest 
(i.e., $\delta \ve{B}=\delta P = \delta \ve{v}=0$). 
The gravitational wave excites only a magnetosonic wave 
when $\ve{B}_0^z=0$. The solution of the perturbations is 
given by analytic expressions in Eqs.~(\ref{case1:solv})--(\ref{case1:solB}). 
When $B_0^z \neq 0$, the gravitational wave excites 
fast- and slow-magnetosonic waves and an Alfv\'en wave. The solution 
in this case is given by Eqs.~(\ref{case2:solv}), (\ref{case2:solP}) 
and~(\ref{case2:solB}).

We have performed numerical simulations for MHD waves 
induced by linear gravitational waves with a new 3+1 GR MHD code 
and have found good agreement with the analytic 
solutions presented here. The details of our numerical code, 
together with a comparison of numerical and analytic solution 
for these waves, as well as the results of other code tests, 
are reported in Paper~I.

\acknowledgments

This work was supported in part by NSF Grants PHY-0205155 and 
PHY-0345151, and NASA Grants NNG04GK54G and NNG046N90H at the University of 
Illinois at Urbana-Champaign.


\begin{thebibliography}{99}
\bibitem{fn0} The first GR MHD code was written by
Wilson~\cite{wilson77} using conformally flat approximation.
\bibitem{ft0} The calculation in~\cite{psvk01} was done in coordinate 
basis. However, the authors studied a special case in which the magnetic 
field is perpendicular to the propagation of the gravitational wave and the 
Alfv\'en speed approaches the speed of light.
\end{thebibliography}
\end{document}